# Model of Trust Management for Digital Industry Services. Towards E-Commerce 4.0

Wolfgang Bauer (1), Natalia Kryvinska (1), Jürgen Dorn (2)
(1) Comenius University in Bratislava, Information Systems Department
(2) Technical University in Vienna, Institute for Information Systems Engineering

## ABSTRACT

The progressive digitalization is changing the way businesses work and interact. Concepts like Internet of Things, Cloud Computing, Industry 4.0, Service 4.0, Smart Production or Smart Cities are based on systems that are linked to the Internet. The online access to the provided data creates potential to optimize processes and cost reductions, but also exposes it to a risk for an inappropriate use. Trust management systems are necessary in terms of data security, but also to assure the trustworthiness of data that is distributed. Fake news in social media is an example for problems with online data that is not trustable. Security and trustworthiness of data are major concerns today. The speed in digitalization makes it even a greater challenge for future research. This article introduces therefore a model of online trust content of service advertisements, which is incorporated into Linked-USDL. It contributes to standardize business service descriptions necessary to realize visions of E-commerce 4.0. It is the basis for building trust enhancing architectures in B2B e-commerce. To do so, we conducted case studies, analysed websites, developed a prototype system and verified it by conducting expert interviews.

## KEYWORDS

E-Commerce 4.0, Business Service Description, Service Digitalization, Trust Management, Trust Signalling, Linked USDL, Service Purchasing, Industry Procurement, E-Markets

## 1. Introduction

"Digitalization" is one of todays' buzz words of Europeans politicians. Its realization is however manifold. One hopes that investments in grid infrastructure enables the implementation of concepts and visions of Industry 4.0, service digitalization, smart cities, etc. and thus to track the main goal to stay competitive on the global market. A market where labour-intensive production already moved from industrial countries to regions with low labour costs like China. High wages in industrial regions are the reason that European countries have to focus on machine intensive productions like for example the assembly of a car, the production of steel or the production of chemicals, where the share of wage costs compared to total product costs is low. Concepts and visions of Industry 4.0 and Smart Production are important research topics for those economies to stay competitive in the remaining industry fields. Furthermore, the importance of service provision increased in those economies because manufacturing companies add service offerings to their product portfolio to differentiate from competitors and to provide customized solutions rather than products. Another reason for the increase service orientation is the outsourcing of activities in the industry domain. This trend towards services being increasingly supplied together with physical products is a phenomenon also described as the 'servitization of manufacturing'. Industrialized countries are undergoing a structural change towards a service society. The service sector is growing, however the productivity in this sector is typically much lower compared with the industry sector and a lack of research in services exists (Maglio *et al.*, 2009). Service Science emerged which tries to understand services to improve the efficiency of creating new services and to improve the productivity in this sector (Dorn and Seiringer, 2013). Industrial Product Service Systems (IPSS) is one specialism of Service Science, due to the importance of the combination of products with services. Service research is an essential field for industry economies where buzz words like Service 4.0 are used to describe the digitalization of services.

The globalization is a main driver that forces industry economies to change. Especially, changed political frameworks, like the enlargement of the EU, the end of the "Iron Curtain" to the East or phenomena like the industry race to the Chinese market at around 2000, are some causes of today's globalized market. The emergence of the World Wide Web has supported this trend, because it increased the transparency of information of goods and services available and thus, it increased the reach of sellers and buyers and as a result it provided an improvement of the matching of buyers and sellers. E-commerce systems were developed to support the trade transaction processes using web-technologies. As the 4.0 hype does not stop at E-commerce, solutions by Artificial Intelligence are driving this agenda like intelligent product and service recommendations or end-consumer communication via AI. A challenge that faces E-commerce 4.0 in the B2B domain is to interface the Smart Factories where intelligent systems make requests that will be handled by automated agents that execute orders for manufacturing and purchasing parts. Remaining competitive in a high-wage society, of course, requires such optimizations to higher efficiency in the supply chain. However, the main challenge for E-commerce systems is the increased service orientation in the industry. Heterogenic business service requests and advertisements needs to be matched on a global market. This is not only a concern of supply chain activities; it is even more important for the marketing departments.

Nevertheless, in the B2B service domain, that visions that automated agents match service requests and advertisements are far from the reality, mainly because todays' industry solutions are highly heterogenic and technical complex product-services rather than simple products. Furthermore, customers and of course automatic agents cannot check the outcome and quality of an offered solution because the service part represents an intangible performance, which cannot be used and verified before consumption, e.g. an engineering service. Bauer and Dorn (2016) described in that regard the delivery of a casting part in the automotive domain, which requires a long development process including a technical consulting to choose the best manufacturing technology, the manufacturing of casting tools, the delivery of prototypes and several product quality optimization cycles in accordance with the customer wishes. At the end, maybe the customer must change the technical specifications of the part, when the efforts show that some of the whished dimensions are not feasible for manufacturing with the chosen technology. Services in the industry include domain specific business transaction processes like quality and contract management to reduce risks, because of uncertainties of service outcomes. Therefore, trust evaluation and trust management are integral parts of every business service interaction. In fact, the E-commerce support for such industry service transactions is at a low stage of maturity. Most frequently, only homepages of companies are used to advertise those description-complex and individual business services (Pedrinaci, Cardoso and Leidig, 2014), which is known as E-commerce 1.0. Manual Google search is therefore the standard process to acquire information before interaction starts and quotations are requested. Often, companies use Excel or custom-made database solutions to support the inquiry processes and the comparison of offers.

The challenge towards a realization of E-commerce 4.0 is to standardize the description of those business service offerings and processes to support the main stages of the procurement process like search, discovery, comparison, and negotiation of services by electronic systems. Such standardizations are an interdisciplinary research task which require comprehensive effort of academics from different fields, like business, information systems and product/service dependent engineering areas, but also service marketing or service science and other studies can be relevant. Business service know-how must be formalized in a way that computer systems can access it, so that the visions of AI in e-commerce can be realized. Using formal descriptions like RDF and Linked-Data principles could be a solution to make business know-how available for information systems engineering. This article, therefore, demonstrates an enhancement of existing service description concepts. It provides a formal model of trust assertion as part of the service interactions process. Trust management systems must be an integral part of E-commerce 4.0 visions because of the increased perceived risks at service business. Those intelligent systems need to access online content to evaluate the trustworthiness of a provider and thus the online service offer as basis of an intelligent service recommendation.

## 2. Problem Description

Researcher in Supply Chain Management discovered, that service purchasing is perceived more risky than goods procurement, due to the implications of intangibility and heterogeneity, which causes uncertainties (Perez-Cabanero, 2008). Especially in industrial environments, inappropriate service performance could result in high follow-up costs. Service marketing explains those increased perceived risks in service trading by the information asymmetry between buyers and sellers, that creates strong incentives for sellers to cheat on services. The best example therefore is a car repair service, where the mechanic is the expert and the consumer the layman. Different kind of frauds can be labelled in such a situation of information asymmetry as overtreatment (charging for a filter that is not required), overcharging (charging for a filter that is not changed) or under treatment (don't change the filter, even if it is required) (Dulleck and Kerschbamer, 2006). Consensus is present within the service and relationship marketing literature that trust is required whenever risk, uncertainty, or interdependence exists (Mayer, Davis and Schoorman, 1995). Trust is important for business interactions to overcome these risks and to engage in assistive actions in environments characterized by uncertainty (Luhmann, 1988). The evaluation of trust signals is therefore a permanent process in service purchasing to reduce the perceived risks caused by uncertainties of services. Signalling is a strategy of sellers in situations with information asymmetry. Signalling theory is applied to scenarios that occur in a range of disciplines not restricted to micro economy (Bliege Bird and Smith, 2005). In the case of a car seller, signalling would be the offering of an inspection by a certified garage to provide additional information, showing that the offered car is good. A researcher signals his abilities by showing the list of publications in high graded journals. These additional signals need to be described at service advertisements by electronic systems. Homepages are used by marketing experts as their major medium to promote their services and to signal their trustworthiness online. So, customers assert their subjective trust values based on unstructured information content that is publicly available. Based on a provider's known trustworthiness, also the unknown service is trusted (Aljazzaf, Perry and Capretz, 2011). An additional problem in online environments is, that e-commerce is more impersonal compared with traditional commerce, more automated, provides fewer direct sensory cues, has less immediate gratification, entails more legal uncertainties and presents more opportunities for fraud and abuse (Head, 2002). So, the digitalization of the service industry requires trust management systems. Service description languages need to incorporate this trust assertion process as part of the service interaction description, including the structured content of information that is considered as trust signals by business experts. Intelligent agents must be able to check the trustworthiness of an online service offer by those signals.

## 3. Related Work

Regarding service discovery and selection, a great amount of work is being carried out on Internet service description and standardization, especially in the area of Service-oriented Architecture (SOA) and Web services. Examples of work in this directions are WSDL/UDDI, SoaML, OWL-S, WSDL-S, SAWSDL (Sun, Dong and Ashraf, 2012). Although (semantic) Web services work provides advanced support for discovering or composing technical services, it disregards the fundamental socio-economic context of real-world services (e.g., value chains and offerings), and does not cover the widespread manual services (e.g., engineering) (Akkermans et al., 2004). They describe services from pure technical point of view, which is not sufficient for the specification of business services delivered or even advertised via the Internet. The term business service is used differently between academics of business-related fields and researchers of information systems. While informatics people consider business services as software packages providing business functionalities (Benfenatki et al., 2017) this research applies the broader view of the business academics, who see a business service as an abstraction of a business process (Jennings et al., 2000; O'Sullivan, Edmond and Ter, 2002; Schuster et al., 2000), usually a human process possibly supported by different technologies including IT. In business perspective, a service is a process that transforms input of the customer and the provider to an individualized output, which could be many different things. Thus, co-creation, pricing, legal aspects, and security issues are all elements which must also be part of service descriptions (Cardoso and Pedrinaci, 2015), as well as the description of the trustworthiness of the service and the provider (Bauer and Dorn, 2016; Bauer and Dorn, 2017a, 2017b, 2017c, 2018). Literature surveys classify the efforts in service description languages (Benfenatki et al., 2017; Kadner et al., 2011; Sun, Dong and Ashraf, 2012) and clearly show, that the focus is rather on Web service and SOA description, cloud service description and software as a service (SaaS) description as studied by informatics academics. Those languages do describe what a service does with little regard to the environment they are deployed in (Benfenatki et al., 2017).

The identification of the problem that business service description needs to be standardized in order to support transaction processes online is already mentioned around the turn of the millennium (Dorloff, Leukel and Schmitz, 2003). But only a view approaches appeared that formally describe business relevant concepts of services. Those research efforts can be distinguished into three categories of studies. First, research is restricted to a specific business branch e.g.: consulting (Bode, 2019; Greff and Werth, 2017) or in mechanical manufacturing an example is the Manufacturing Service Description Language (MSDL) (Ameri, Urbanovsky and McArthur, 2012). Second, studies focus on the description of one specific business aspects of services like service costs (Seiringer and Bauer, 2016), or this study focusses on the description of trust in services. Third, some studies try to generalize service descriptions for all kind of services. A major contribution in this regard is O'Sullivan (O'Sullivan, 2006) who describes non-functional properties (NFPs) of services, like price, location, time, etc., but trust description remains in this taxonomy in an abstract layer, which limits this research.

USDL is the subsequent research that tries to unify descriptions for all kind of services. Its roots are in funded projects, like the FI-PPP or the THESEUS (Oberle and Barros, 2012). Attensity and SAP Research among others initiated this language that was submitted for standardization to the W3C (Benfenatki et al., 2017). It is the latest comprehensive language which provides a (multi-faceted) description aiming the commercialization of (business and technical) services over the web to enable the trade of business services by the emergence of internet marketplaces (Cardoso et al., 2014). The roots of USDL are already a decade ago, the latest official version is USDL 4.0, which is also called Linked USDL. The advantage of Linked USDL is that it uses the principles of Linked Data, so it is easy extensible and thus researcher are still refining it for different purposes. Furthermore, Linked-USDL makes use of other vocabularies that promote structed data on the internet and focusses on the description of e-commerce data, like schema.org. Compared to USDL or MSDL, schema.org does not explain service processes, it only structures data that is frequently used in e-commerce transactions. The actual version of USDL is divided into different levels that have different maturity. Each module is a set of concepts and properties. The main module USDL-core describes operational aspects of services. USDL-sec cares about security properties and processes. USDL-price describes prices structures, USDL-agreement represents quality issues of services like response time and availability, finally, USDL-ipr formalises the rights to use a service.

Linked USDL is a comprehensive vocabulary for capturing and sharing rich service descriptions, which aims to support the trading of services over the Web in an open, scalable, and highly automated manner (Pedrinaci, Cardoso and Leidig, 2014). It can be used as basis for further development of business descriptions. Linked-USDL does not describe the process of the evaluation of the trustworthiness of the service provider and the required data, which is necessary in service interactions and even more essential when doing business via electronic systems. So, this research will introduce trust assertion as an essential component of business interactions and proposes an extension of the Linked USDL vocabulary to enrich service description.

## 4. Methodology and Research Process

The Design Science Paradigm in information systems research has been followed (Alan R. Hevner, Salvatore T. March, Jinsoo Park, Sudha Ram, 2004) to plan the exploratory research process. This research paradigm represents a framework that was used as a guideline during the complete research study. The design science paradigm defines the "environment" as the problem space, a composition of people, (business) organizations, and their existing or planned technologies. This environment defines the business need or "problem" as perceived by the researcher. The problem must be real and relevant in this environment. It defines research as a process with multiple evaluation cycles where results must be published and incorporated into the current knowledge base. Therefore, several research steps have been conducted and different methods were used. Results were published in numerous papers to assure regular feedback loops. This article will provide an overall summary of the developed trust model that is linked to USDL to improve the knowledge base. Furthermore, the verification of the enhanced service description by expert interviews is presented.

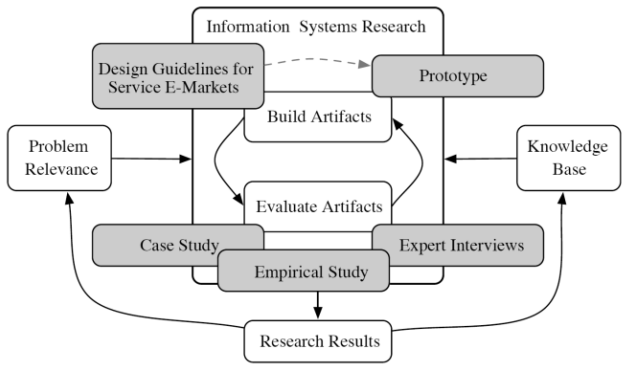

Figure 1: Research Process

Accordingly, the first step was to analyse the requirements in trust descriptions for services in real business environments. Therefore, case studies were conducted in the industry domain. Procurement specialists were interviewed and asked, which information they consider during the service discovery process. Especially, their trust aspects were studied and a conceptual model was developed (Bauer and Dorn, 2016; Bauer and Dorn, 2017a). Such identified aspects are for example, the company size, company's financial health, the company's history, resource descriptions like employees, machines and production facilities, company's location, control mechanism, customer references, product and service portfolio or known experiences of business partners with the company, to mention some examples. However, not all the trust aspects considered by buyers can be made public available at business service advertisements by electronic systems. For example, the customer's experiences with a supplier are usually stored in the customer's supplier evaluation systems to share it among its organization. Those experiences are part of corporations' competitive know-how and therefore a company's secret that is not shared with other market participants and competitors, unlike in B2C situations where ratings of consumers are shared and used as a major trust feature provided by reputation systems in online environments. That is why a second research step focused on trust signal descriptions that are published on websites of service sellers in the B2B domain. The question of: "Which trust data is published by B2B service providers due to signal trustworthiness" was answered by analysing fifty different professional service company websites (Bauer and Dorn, 2018). Those homepages are used by marketing as major medium to promote services and to signal trustworthiness to potential business partners.

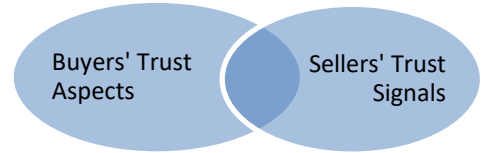

Figure 2: Relevant Trust Content

Although, marketing uses established concepts for advertising their service offers by signalling trust, those websites lack in semantics. So, each site was observed manually, all trust signals were collected and categorized. The resulting analysis of the intersection shows which trust aspects used by purchasers (research step 1) are publicly available (research step 2). Thus, we discovered a set of trust data objects and their properties that are frequently used and published online by marketing to signal trustworthiness and used by buyers to evaluate the trustworthiness of the service provider (Bauer and Dorn, 2018). We call this intersection "**Trust Content**".

The received data was then used to build a formal description of the online trust content accessed during the trust assertion process in the service industry. This model was subsequently linked to Linked-USDL. The Resource Description Framework (RDF) was used as formal ontology representation languages to capture the semantic content of online trust signals of services advertisement and to describe the process of trust evaluation by a buyer as part of the service interaction process. Consequently, the proposed Linked-USDL extension consists of both, the description of the relevant trust content (data) and the process of trust evaluation (Bauer *et al.*, 2019). This model enhances the knowledge base of service description.

Finally, the proposed trust model was implemented by building a prototype e-market system, realized by a student as content of a master's thesis. The developed prototype was simply a kind of e-market system that focused on the representation of the information described by the developed trust model. The prototype was filled with the real data of fifty websites, that where analysed earlier.

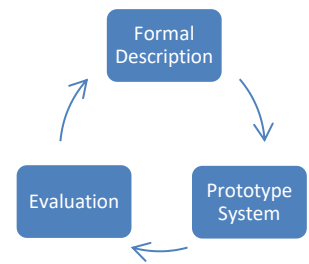

Figure 3: Evaluation Cycle

The prototype was then used during subsequent expert interviews to evaluate the developed model and to discuss with experts, which system features they would propose to support the trust management process. Therefore, the prototype received some dummy "**system trust features**" known from B2C businesses, like a rating functionality and the representation of system data like date of registration, number of profile clicks, number of ratings, etc. Also, an algorithm to calculate a total trust value out of the trust content was implemented as a proposal for discussion. So, an evaluation of the developed model of "trust content" could be performed and experts' opinions to "**system trust features**" could be analysed to propose a model for trust management in e-commerce for services.

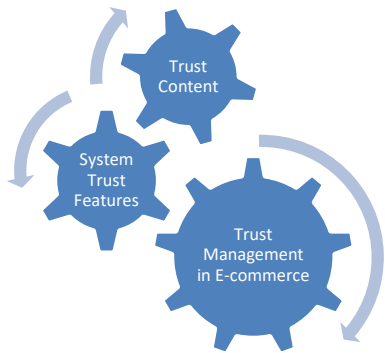

Figure 4: Trust Management in E-commerce for services

A semi-structured interview guideline was used therefore including open and closed questions; however, the interview guideline leaded from closed questions to a more open conversation. Therefore, the MoSCoW rating (must, should, could, won't) served to prioritize the trust content with an open comments' category, which left room for participants to explain their response. The coding of the quantitative data involved assigning a numerical value to each response, e.g. 1=must have and 4=not recommended description, so that statistical values could be computed. 50% of the interviewed persons were managers, who lead in average ten buying specialists. In average the interviewed managers had an experience in their field of 11 years. On the other hand, 50% of the interviewees were procurement specialists with an average experience in buying industrial services for 14 years. In total, ten interviews were conducted. All respondents worked in the field of industrial purchasing at following branches: car manufacturing, automotive systems supplier, intralogistics automation, electrical power supply, metal industry (knife production) and electronic systems development and manufacturing.

## 5. USDL Extension for Trust Assertion (USDL-trust)

The following section describes the extension of USDL by the process of trustworthiness evaluation in business service interactions. As basis concept it makes use of the Trust Assertion Ontology (TAO) (Sacco, 2013). The TAO was developed in the context of Online Social Networks, where personal data is stored, managed, and shared with other people. The TAO ontology describes trust judgments made by users applied to entities that request personal information. TAO describes the subjective measurement of trustworthiness of the requesting entity by the use of different factors like identity-based trust, profile similarity, reputation in trusted networks, relationship-based and interaction-based trust (Sacco, 2015). Like Linked USDL, the TAO ontology is modelled in RDF. Both concepts apply Linked Data principles and make use of external vocabularies. TIME [1], SKOS [2], GOODRELATIONS [3] are used by Linked USDL, the TAO ontology incorporates the FOAF[4]. Linking both concepts requires additional properties and objects that will receive the prefix usdl-Trust, which will be used for "USDL extension for trust assertion". The usdl-Trust makes use of additional ontologies such as SCHEMA.org[5] and DC[6]. In the following the linkage between the TAO ontology and Linked-USDL will be introduced.

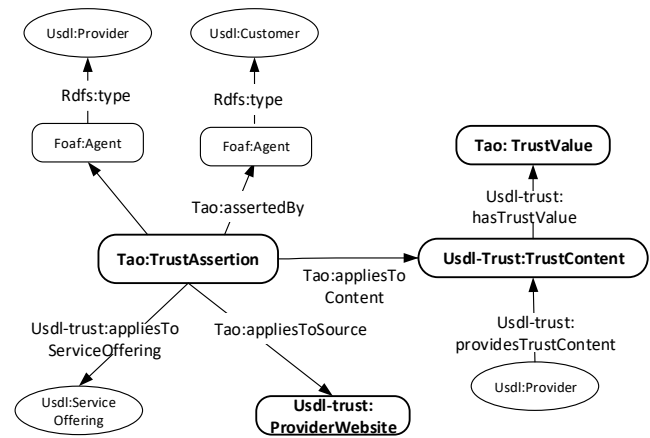

Figure 5: Incorporation of TAO into Linked-USDL

The TAO ontology is created around the main subject Tao:TrustAssertion, which is the process of a customer who asserts a subjective trust value to a provider based on different contents, which are trust signals provided by the provider. The tao:TrustAssertion applies to an foaf:Agent and is asserted by another foaf:Agent. Here the linkage to USDL is obvious the usdl:Customer who asserts trust to the usdl:Provider, which are types of foaf:Agent. Although, trust evaluation in business environments applies to other information contents than in social networks, the process of trust assertion is the same, information content of a specific source is evaluated, and a subjective trust value is asserted to each trust aspect. The trust assertion in business service interactions applies to the content of usdl-Trust:TrustContent. Perceived trustworthiness is subjective and varies between people, so the influence of certain information on the evaluation of trustworthiness varies depending on the people (Bauer and Dorn, 2016), so each customer asserts an individual trust value to a provider based on the provided content.

---

[1] https://www.w3.org/TR/owl-time/

[2] https://www.w3.org/TR/2008/WD-skos-reference-20080829/skos.html

[3] http://www.heppnetz.de/ontologies/goodrelations/v1.html

[4] http://xmlns.com/foaf/spec/

[5] https://schema.org

[6] http://www.dublincore.org/specifications/dublin-core/#current

The TAO provides two properties of the subject tao:TrustAssertion that makes the concept universal applicable for different domains: tao:appliesToSource and tao:appliesToContent. In online business contexts trust assertion is made by the use of trust signals published on provider's websites. Therefore, the definition of a new object, the usdl-Trust:ProviderWebsite is necessary. Trust assertion is a process that applies to every service offering in different intensity, depending on factors like customer risk perception, customer's experiences, or personal beliefs.

This simple extension of the Linked USDL vocabulary does add an important business perspective to service interactions. It makes Linked USDL completer and more accurate to apply it in B2B environments. But, supporting the trust assertion process by information systems requires to structure the content used by people in a machine-readable format. Although, trust assertion always is a subjective process that cannot be resolved automatically by web-services, information technology must support this process. Therefore, a formal description of usdl-Trust:TrustContent is needed. The next section will structure the discovered content used to assert trust in service businesses.

## 6. Model Development and Verification

This section demonstrates the development of a trust model that formally describes trust content that was identified by analysing service websites to answer the question: "Which trust content is used by service providers to signal trustworthiness?" Furthermore, it introduces the opinions of the experts and their prioritizations at trustworthiness assessments. It indicates, which data objects must be described by systems that support service interactions and enable users to assess their trust in service advertisements. Eight categorizes of trust content were identified: Customer Reference, Certification, Publication, Facility, Employee, Partner, Legal Data and Terms.

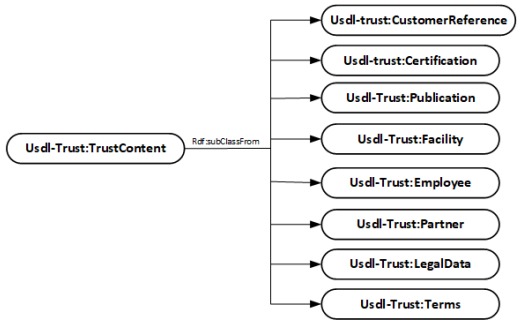

Figure 6: usdl-Trust:TrustContent

### 6.1 usdl-Trust:CustomerReference

Customer references are signals of provider experience. Three main kind of references were discovered on the examined websites, that provide a direct link to a customer. About fifty percent of all examined websites show selected logos of their customers, another fifty percent of the analysed websites list customer names as a reference. Also, projects descriptions or written testimonials of satisfied customers were observed. In total, 76% of all service companies signal trustworthiness by displaying any information of customers (Bauer and Dorn, 2018).

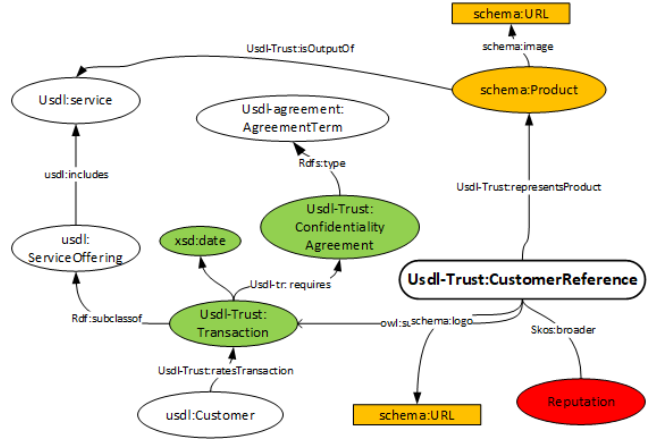

Figure 7: Usdl-Trust: CustomerReference

Linking usdl-Trust:CustomerReference to USDL requires the introduction of an additional object the usdl-Trust:Transaction. This object is a subclass of usdl:ServiceOffering, but receives a transaction date, which is the date of the contract. The introduction of the usdl-Trust:Transaction allows an usdl:Customer to rate a transaction, which is then also available as customer reference. Customer references often represent parts produced for a customer, so they represent schema:Product, which is an output of an usdl:Service. Often a picture of parts are provided, e.g.: a prototype that was manufactured. Also, a textual description of such tangible outcome is used to provide a business reference.

Experts rated the importance of the description of customer references at online service advertisements as 1.6 (1= must be described, 2= should be described). They commented that *the logo of the customer catches someone's eyes* (schema:logo) and that *the customer name is supportive* (usdl:Customer). Furthermore, they added, that the suitability of a picture of the tangible outcome (schema:Product) *depends on the product or service itself, e.g. at plastic parts, an image is an important description. On the other side a printed circuit board (PCB) always looks the same, a picture does not work.* Another respondent added that *a picture would be fine and text descriptions as further information after clicking on the picture would be suitable.* A buyer from the automotive branch mentioned that, *text descriptions are difficult, often such information is not allowed, because of non-discloser agreements. A textual detailed description would be very important, but in the automotive domain seldom possible,* which was the reason to add the usdl-trust:ConfidentialityAgreement to the model, which is necessary to allow a system and/or a provider to filter the transaction data, that is allowed to be published as customer reference for signalling trust. The confidentiality agreement is interfacing usdl-agreement because it is a type of an agreement term.

## 6.2 usdl-Trust: Certification

Certificates are used to signal trustworthiness by the use of an independent third-party referee that audits specific capabilities of a company. Used certifications in the industry service domain are for example the quality standards ISO 9001, ISO14001 and the ISO16949. More than 90% of the analysed companies display such certifications on their website by providing a short textual description or a hyperlink to a copy of the certificate (Bauer and Dorn, 2018).

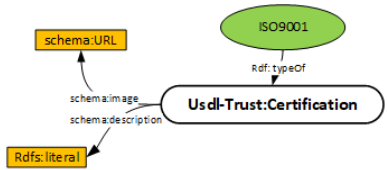

Figure 8: Usdl-Trust:Certification

In average the interviewed experts value the importance of describing certifications to evaluate the trust in the promised service as 1.8. It exists different opinions about the implementation of this description like one respondent commented: *Certain certifications must be described, like the ISO9001, whatever text or picture* another respondent said: *As former quality manager, I want to see the certificate as a picture.*

## 6.3 usdl-Trust: Facility

Business companies can possess different facilities like production facilities, sales facilities, a head office or a holding for example. All companies share data about their facilities, at least the address of their main location. USDL uses the GOODRELATIONS vocabulary (gr:Location) do describe those facilities, where certain trust related key data is described, like the address of the facility. An usdl:Service-offering is available from a gr:Location.

Many companies provide pictures of their facilities, even some use a google maps plugin on their website to show their facility from the top, including neighbours or enabling street view of the facility buildings to provide physical evidence of their facility. For the description of the picture of the facility a foaf:image is linked to usdl-Trust:Facility by the property usdl-trust: hasImage. Also, historical data like the foundation year of the facility and the number of employees at the facility are published on websites of service providers. Therefore, we added to usdl-Trust:Facility another property usdl-Trust:hasKPI that points to an object usdl-trust:KPI.

Experts valued the importance of facility descriptions in average as 1.8 (should be described). Some variations can be explained by the different branches the respondents come from, like one expert from electronics manufacturing commented: *The location is decisive, but not how many locations the company has, e.g.: sometimes (in the case of more complex parts) it is necessary to find a company that has local headquarters (Europe) with global production (Asia), providing a close contact in Europe to handle problems, but delivering directly from their Asian facility to our Asian subsidiary. In the case of more simple parts, no European headquarter is needed.* Another respondent from the automotive industry added: *Beside address, picture and contact to the facility, some KPI of those facilities would be interesting, such as the size of the production area. In addition to no. of employees or annual turnover, the yearly investments are indicating health and prosperity of the company.* Another person from a car manufacturing company said that, *the annual turnover does not give us a correct indication for the health of the supplier; our suppliers must provide in addition detailed financial data including profit and loss calculation. I know that this is specific for our branch.* Furthermore, an expert supplemented that, *the description of organizational and group structure is important,* so the usdl-Trust:Facility needs to be linked to the schema ontology by usdl-Trust:Facility -> usdl-Trust:belongsToOrganization -> schema:Organization.

The quantitative data analysis disclosed that professional marketers do publish additional content to describe trustworthiness by their facilities like used machines. To describe this additional facility data an object usdl-Trust:System is created. It is a subclass of gr: Locations.

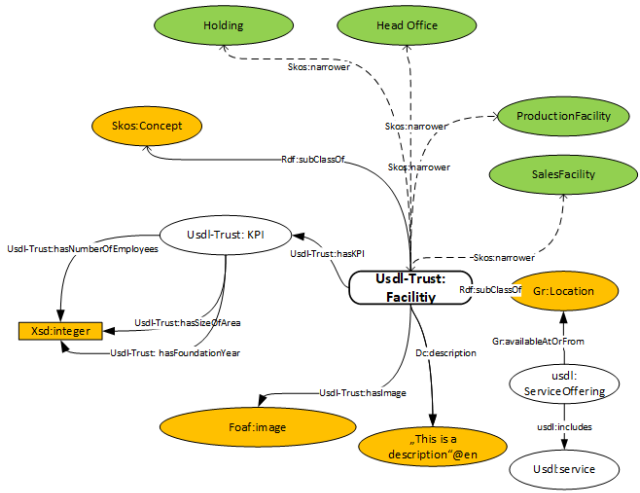

Figure 9:Usdl-Trust:Facility

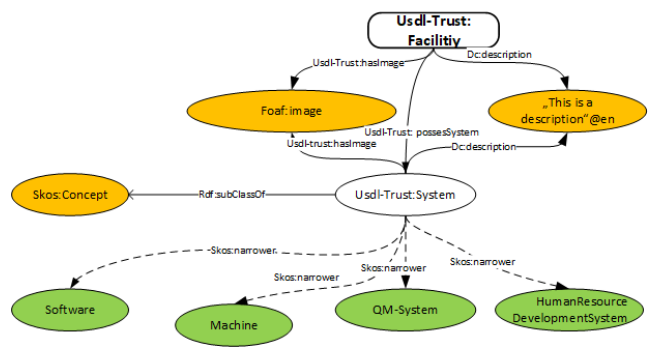

Figure 10:Usdl-Trust:Facility

Displaying systems that are needed as input for service production is another method to signal that the promised services can be carried out by the offering company. Marketing claims that this physical evidence increases trust in services (Shanker, 2002). Noticeable was, that manufacturing service providers, almost always promote their services and trust by presenting systems, like their machines, their used software systems, their quality systems or other organizational systems, but intangible service providers like e.g.: consulting companies, do rarely display their support systems. So, in total only one third of all analysed websites presented their systems for signaling their abilities. The usdl-Trust:facility possesses this usdl-Trust:system in the proposed usdl-Trust extension module.

Experts value the description of systems in average with 2.0 (should be described). Different suggestions for the implementation of those descriptions were provided by the experts: *Machinery park description is important by pictures including text descriptions; Pictures of machines, how they are installed in the production facility makes a better impression, than only a picture of the machine from a sales catalogue; Having machines of recognized machine builders of e.g. Europe makes better impression than owning cheap Asian machines, so the name of the machine producer is a relevant information.* Also, experts gave explanations why those descriptions are of importance like: *Machine descriptions are an indication for the vertical integration, showing, if someone is producing or buying the necessary manufacturing parts. The more vertical integration of the company the more know how exists.* Experts of the industry domain commented mainly to machine descriptions, but also referred to other systems: *The used ERP system is interesting to understand, if an EDI interface could be installed,* or: *Detailed information about how they organize their production like Kanban (pull) vs. MRPII (push) system is too detailed and not of interest for the buyer.* We conclude that the required systems for service production are important trust signals, but priorities depend on the service branch.

### 6.4 usdl-Trust: Employee

Services are as good as its people (Berry and Parasuraman, 1991), therefore service provider make online content available about their employees to signal ability. The analysis of websites discovered, that 85% of all websites display information of their service personnel. Following data is used to describe employees, which can be used from the schema.org vocabulary: name, job title, honorific prefix, email, telephone number, picture of the person and a description what the person knows about (expertise). Compared to the eight employee properties provided on the observed websites, the schema.org description provides 57 different properties for schema:Person. Usdl-trust:employee can be seen as an instantiation.

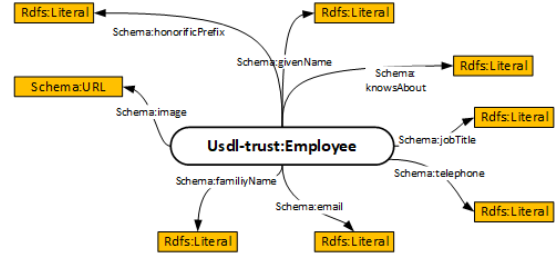

Figure 11: Usdl-Trust:Employee

Experts value the importance of the description of the employee as 1.4 (=must be described). They see it as a *big hassle* if contact persons are not described at service advertisements or they say, *nothing is worse than only providing a contact form on the website, without direct contact to the persons of interest.* An expert commented: *The name, contact and job position are the most important data, after the picture of the person. The CV is only of interest, when the owner of a company or upper management is described.* During an open discussion to the description of the employees, another expert added: *The human resource is frequently a topic. Not only the description of the contact person itself is of interest, but how the company handles the problem, that human resources are the most important, most expensive resources and at the same time the tightest.*

### 6.5 usdl-Trust: Partner

The outsourcing of activities and concentration to core competencies in the industry domain increased the importance of supply chain. Therefore, a factor of success is to have the right (best) partners. So, buyers look if providers have supply chain expertise by checking signs of their partners. To formalize those information we establish the object usdl-Trust:Partner. Links to company's social networks were identified and classified as a way marketer demonstrate their business network, which is described in usdl-Trust:Partner by the property socialNetwork. It was furthermore identified that partners are described on websites by a textual description and/or their logo.

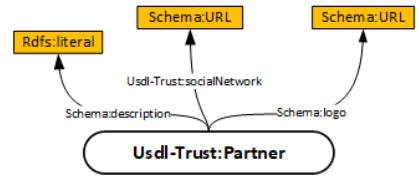

Figure 12: Usdl-Trust:Partner

Experts value the importance to describe partners in average as 2.0 (=should be described). But in this category different perceptions could be identified like: *If a caster uses a standard partner, for example for vertical processing like machining of the raw part, it is an important thing to know, which should be described on the website.* Another respondent said: *Usually, we are not interested in the sub-supplier of the provider.* Digital marketing and marketing in social networks is very important nowadays. However, the interviewed experts do not have interest in information provided by social networks. They don't follow a possible supplier or even check their social media channels. They commented: *Social media information is absolutely unimportant,* or *social media are pseudo friends, this information will not be evaluated in industrial environments.* This seems to be specific for the manufacturing domain. Experts stated that they like to know about strategic partners of the provider, when those partners have a significance in the provider's value creation process.

### 6.6 usdl-Trust: LegalData

Many companies provide legal data on their websites such as their VAT number, their company registration number (CRN), legal identification numberings like the LEICODE or a DUNS code to enable a customer to check identity of the service offeror. Furthermore, different licenses they possess can be part of their trust promotion strategy.

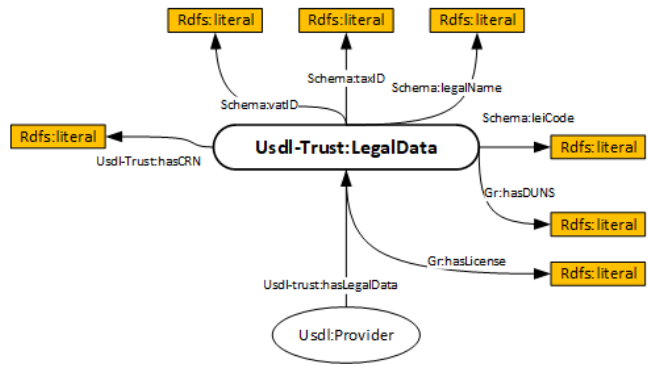

Figure 13:Usdl-Trust:Provider

Experts value legal information of the service advertiser in average as 1.2 (must be described). Different prioritizations of it's implementation could be observed like one said: *Primary, the VAT no. is of interest* another expert commented: *The company registration no. is required for credit assessment to evaluate supplier risks* or: L*egal status and transparency of corporate integration is important,* and provided explanations like: *The legal status is of interest, because we prefer supplier, who are leaded by owners, so that the management has influence on determining decisions.*

### 6.7 usdl-Trust:Terms

Different standard terms are published on websites, like for example General Terms, Terms of Delivery, General Purchasing Terms, General Sales Terms, etc. which provide a kind of control mechanism that is a trust signal (Bauer and Dorn, 2017a). Furthermore, company's policy is kind of trust content that was categorized to terms because those provide an internal directive for the company.

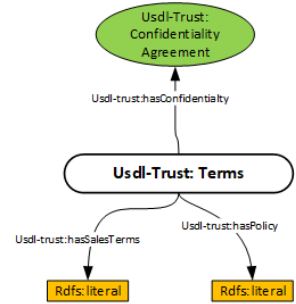

Figure 14:Usdl-Trust:Terms

Experts perceive the description of terms as less important (average value 2.8) and explained it as following: *I am not interested in the terms of the supplier. The supplier has to accept my terms; Displaying terms in general makes a professional impression, but I don't read them in this stage; Sometimes policies are interesting, if you check a company in far east, for example in India. If someone has the policy to give benefits to employees with children, this could be a good sign for less employee fluctuation, which is there a big problem.*

### 6.8 usdl-Trust: Publication

*"Look here at this newspaper, this company has implemented Industry 4.0."* This was a statement of a procurement leader recorded during a case study, showing, that companies do advertise themselves in different media. The company websites are used to provide links to such publications. Success stories, company events and research papers were identified as publications on websites, or as links on websites to signal expertise and ability which is a factor of trust, like integrity (Mayer, Davis and Schoorman, 1995). Therefore, we identified, that companies provide stories of the employees' e.g. their participation in a sports events. In total about 70% of all websites provide different kind of publications to signal trustworthiness. Those contents will be formalised by the use of the object Usdl-Trust:Publication which is a type of schema:CreativeWork. The schema vocabulary is suitable to describe the identified contents.

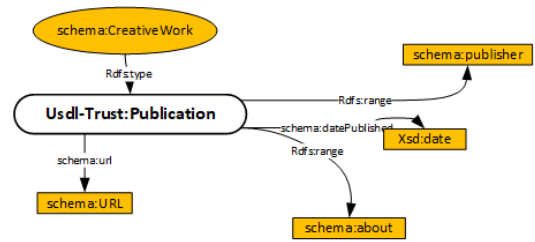

Figure 15:Usdl-Trust:Publication

Descriptions of publications are valued in average as 2.4 by experts. It value and the comments show a lower importance. Particularly, the kind of publication which frequently are published on websites like company newsfeeds aren't needed: *Internal newsfeed does not*

*have expressiveness; Newsfeed is not a trustable source.* Only publications of professional sources are of interest.

## 7. usdl-Trust: System

E-commerce systems serve as intermediary between buyers and sellers (Bakos, 1991) to exchange information about product and service offerings and to provide intermediary services to support the transaction processes. Those intermediaries can also contribute to the process of trustworthiness assessment during the service discovery. So, usdl-trust:System is added to the model, which is providing usdl:TrustContent like data and features.

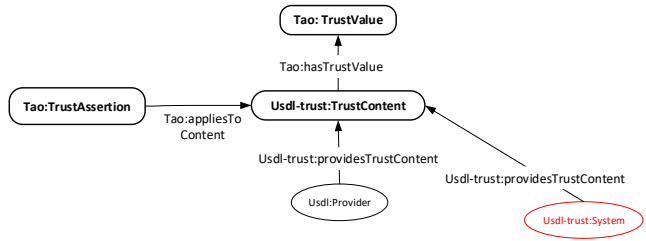

Figure 16: Usdl-trust:System

Such content can be analytical data like number of profile clicks, number of transactions or the date of joining the e-marketplace. Expert valued such data with 2.5 (could be described). Likewise, the intermediary system can support trust by providing warranties or specific policies e.g. return policies. Other trust enabling systems features are live chats, third-party seals, allowance of credit card payments or order status tracking (Mavlanova, Benbunan-Fich and Koufaris, 2012). Experts explained that such features are more typical for B2C product marketplaces than for B2B services, because *in B2B business we don't pay with credit card, we demand invoices with due date 30-90 days; Our services are unique, we cannot return a motor development, like people return shoes in Zalando; individual contracts are used; Third party seals for the website are irrelevant.* A main trust enabling feature of common e-commerce systems is however a reputation system, where users rate transactions. Nevertheless, most experts expressed scepticism towards ratings, they don't trust ratings by other market participants, unless it exists a transparent mechanism by verifying contracts by independent third parties. They say: *Online ratings are not common in professional environments; Ratings could always be faked, especially in low cost countries those are predestined for fake;* A suggested strategy to increase the trust in a rating is to show the profile of the rater. One respondent said, *if there would be a trustable rating, it would be a "must have" on e-markets*, however he cannot imagine a trustable rating functionality in professional industry environments. Companies wouldn't provide their buying know-how to competing companies and publish it. But one third of the respondents didn't comment on this question and rated this feature as a "should have". Another respondent added that *the precondition to trust a rating is to trust the rater's identity,* he supplemented that the marketplace as a mediator needs to ensure the correctness of user's data such as the identity, otherwise the marketplace is not trustable. The open discussion leaded to the difficulty of the identity verification in global online environments.

Although, data like the VAT number can be easily be checked by a web service that provides corresponding name and address of the company that belongs to the number, that does not guarantee, that the person who pretends to possess this identity, it really is. A kind of two-factor authorization would be necessary, like sending a postal letter to the postal address consigned to the VAT number with a PIN registration. One expert proposed that a financial check of a credit bureau is a service that can be used to check the opponent's data. It turned out, that a trustable identification of the users will be a determining factor for the trustworthiness of the data and the marketplace. Another source for trust is "computational trust", when a system uses algorithm that allow to calculate a trust values according to the data provided. Such a feature was implemented into the prototype [10] and experts valued it with 2.0 (should) be implemented. They commented as following: *In general, it is conceivably that the computation of trust as shown in the prototype could beneficial; I think that an algorithm is feasible; If a system can be developed in this direction as shown, it would be a must have.*

## 8. Conclusions

This article presents a model that enhances the knowledgebase in trust assessment for online service advertisements. It is an important building block for the development of trust management systems in B2B e-commerce. The provided trust model was prototypical implemented and verified by expert interviews. It extends Linked-USDL, based on real case studies, a quantitative analysis of service provider websites and buying experts' opinions. It enhances service description, which is the basis for the development of E-commerce 4.0 aiming to apply AI solutions also for the industry domain. But on this way, much further research is suggested in service description e.g. in formalizing business contract terms, quality management processes and transaction processes. Finally, the biggest challenge is, to formalize the functional descriptions of business services. We recommend research in the categorization of service offers as a first step. Descriptions must be linked together as shown in this approach. Someone could for example study the field of engineering services, structure the content in RDF and link it to USDL as well.

The discussions with experts confirm the importance of the use of the trust content provided in the model, although the evaluation of trust is a subjective process and therefore different prioritizations of trust signals could be observed. The progressive digitalization of economies will increase the need for trust management systems to evaluate the trust in online data also in many other areas of research like IoT, smart cities, social media and of course industry 4.0.